\documentclass[usegraphicx]{mn2e}
\usepackage{amssymb, amsmath}
\usepackage{ifthen}

\def \version {clean}

\ifthenelse{\equal{\version}{annotated}}
{
	\newcommand{\comment}[1]{{\parbox{\columnwidth}{[#1]}}}
}{
	\newcommand{\comment}[1]{}		
}

\def\br {{\bf r}}
\def\bv {{\bf v}}

\def\bg {{\bf g}}

\def\bn {{\bf n}}
\def\bk {{\bf k}}

\def \nbar {{\overline n}}

\def \zbar {{\overline z}}

\def \rbar {{\overline r}}
\def \zbar {{\overline z}}
\def \thetabar {{\overline \theta}}
\def \chibar {{\overline \chi}}

\def \bchi {{\boldsymbol{\chi}}}

\def \S {{\rm S}}
\def \O {{\rm O}}

\begin{document}

\title[Luminosity distance perturbations from peculiar motions]{On the perturbation of the luminosity distance by peculiar motions.}
\author[Kaiser and Hudson]{Nick Kaiser$^1$ \& Michael J. Hudson$^2$ \\
$^1$Institute for Astronomy, University of Hawaii \\
$^2$Department of Physics and Astronomy, University of Waterloo}
\maketitle 

\begin{abstract} 
 We consider some aspects of the perturbation to the luminosity
distance $d(z)$ that are of relevance for SN1a cosmology and for
future peculiar velocity surveys at non-negligible redshifts.

1) Previous work has shown that the correction to the lowest order
perturbation $\delta d / d = -\delta v / c z$ has the peculiar
characteristic that it appears to depend on the absolute state of
motion of sources, rather than on their motion relative to that of the
observer.  The resolution of this apparent violation of the
equivalence principle is that it is necessary to allow for evolution
of the velocities with time, and also, when considering perturbations on the
scale of the observer-source separation, to include the
gravitational redshift effect.  We provide an expression for $\delta d
/ d$ that provides a physically consistent way to measure peculiar
velocities and determine their impact for SN1a cosmology.

2) We then calculate the perturbation to the redshift as a function of
source flux density, which has been proposed as an alternative probe
of large-scale motions.  We show how the inclusion of surface
brightness modulation modifies the relation between $\delta z(m)$ and
the peculiar velocity, and that, while the noise properties of this
method might appear promising, the velocity signal is swamped by the
effect of galaxy clustering for most scales of interest.

3) We show how, in linear theory, peculiar velocity measurements are
biased downwards by the effect of smaller scale motions or by
measurement errors (such as in photometric redshifts).  Our results
nicely explain the effects seen in simulations by Koda et al.\ 2013.

We critically examine the prospects for extending peculiar velocity studies to 
larger scales with near-term future surveys.  
\end{abstract}


\section{Introduction}

The perturbation to the luminosity distance $d(z)$ caused by 
cosmological structure is of interest for two reasons; on one hand it can degrade the precision of cosmological
parameters inferred from the Hubble diagram obtained from standard candles such as SN1a
(e.g.\ Riess et al.\ 1998; Perlmutter et al.\ 1999) 
while on the other it can provide a useful probe of structure.

At very low redshift the distance is, in the absence of structure, simply linearly proportional to the redshift and
the dominant effect of density perturbations comes from the
peculiar velocities associated with the growth of structure.  These velocities cause a change to the
redshift through the first order non-relativistic Doppler effect with the result that
that if a source has a positive radial peculiar velocity $\delta v$ then its distance is lower than for an object of the same
redshift in an unperturbed universe; i,e.\ it has a peculiar displacement $\delta d = - \delta v / H$ 
and a corresponding fractional apparent luminosity enhancement of $2 \delta v / H r$.

Exploiting these measurable perturbations to the luminosity as a function of redshift as a probe of structure
has a long and rich history.  In the past these have been limited to very low redshift and in 
such studies it is legitimate to adopt a Newtonian analysis and consider only the effect of the
motions -- the effects here being the same for motions associated with the growth of structure or
for hypothetical motions of sources relative to the cosmic frame that are not associated with
density perturbations (we shall refer to such motions as ``unsupported'').  But to understand
the effect on supernova cosmology, or to exploit motions as a probe of structure using data from
deeper future surveys, a more detailed treatment is needed that allows for the non-negligible 
redshift of the sources.  For one thing, one cannot simply apply the non-relativistic velocity 
addition formula to relate velocity and distance perturbations.  And for finite redshifts other
effects come into play.  These include the gravitational redshift (Sachs \& Wolfe 1967);
the effect of aberration which modifies both the angular size and also the surface brightness
of sources; and the effect of weak gravitational lensing, an effect distinct from aberration, which
modifies sizes of galaxies leaving their surface brightness unaffected.  The purpose of this paper is to 
elucidate some features of the observable effects of peculiar velocities at non-negligible redshifts.

The structure of the paper is as follows:
In the following section (\S\ref{sec:review}) we provide a review of previous work.  We briefly describe the history
of low-redshift peculiar velocity observational studies (\S\ref{sec:review-standard-candles}) and we also contrast these with other probes
of cosmic motions such as the kinematic SZ effect (\S\ref{sec:review-ksz}); the observer motion induced dipole anisotropy of
isotropic background radiation fields and the dipole in the source counts of very distant
galaxies (\S\ref{sec:review-dipoles}); the effect of motions at low-redshift on structures in redshift space 
(\S\ref{sec:review-zspacedistortion}) -- sometimes
called the `rocket-effect'; and the weak lensing effect (\S\ref{sec:review-gravlensing}).  
In \S\ref{sec:review-theory} we review recent 
theoretical developments.  These are mostly based on the pioneering work of Sasaki and
co-workers who, starting in 1987, provided a framework for describing
the effects of linear density perturbations on the luminosity distance that includes the
effect of peculiar velocities, gravitational lensing and the gravitational redshift (including the
integrated Sachs-Wolfe effect).  Many of the more recent studies isolated the terms that
depend only on the peculiar motions and have applied the resulting simplified formulae
to supernovae cosmology and to make forecasts for how future surveys can provide constraints
on large-scale structure (coining the terms `Doppler-lensing' and `anti-lensing' to describe
these effects, though, as we note, these are not particularly novel probes).  We show, however,
that the modifications to the normal Newtonian analysis that become non-negligible at
finite redshift have a puzzling feature in that they appear to depend on the absolute
motion of the sources relative to the cosmic frame.

In \S\ref{sec:deltamofz} we address the apparent violation of the equivalence principle in the above studies.
We show how these apparent violations are banished once one allows for the decay of the
peculiar velocity in the case of unsupported motions and, in the case of motions
associated with structure, once one includes the Sachs-Wolfe effect.

In \S\ref{sec:deltazofm} we consider a novel proposal to measure peculiar velocities which is to
try to measure the perturbation to the mean redshift as a function of flux density (rather
than the perturbation to the flux-density as a function of redshift).
We show that there is a non-trivial observable influence from the modulation
of the surface brightness in this method.  But we argue that there is a problem
with this method in that, for most scales of interest, the observable is swamped
by the effect of galaxy clustering.  We then conclude with a discussion (\S\ref{sec:discussion}) including
a critical analysis of the prospects for extending peculiar velocity studies
to larger scales with future surveys, and in appendix \ref{sec:appendix} we show how
measurements of large-scale motions are biased by smaller scale structure.

\section{Probes of peculiar velocities and distances}
\label{sec:review}

\subsection{Distance perturbation from standard candles}
\label{sec:review-standard-candles}

Exploiting $\delta d / d$ as a probe of structure -- so called `peculiar velocity' or `cosmic-flow' studies -- has a long and rich history.
Rubin et al (1976) found a dipole moment of flux densities of galaxies with $cz$ in the range 3,500-6,500 km/s,
indicating a motion of the Sun with respect to this shell at a speed of $\simeq 600 $km/s and also a
substantial motion of galaxies in this shell with respect to the cosmic microwave background frame.
This was a most surprising and unexpected result since, outside of the local group,
the Hubble flow appeared to be remarkably cold and close to linear.
In a similar study, but with a more careful treatment of biases from selection effects,
Tammann, Yahil \& Sandage (1979) compared magnitudes of galaxies in the Virgo cluster with 
those of galaxies at the same redshift scattered around the sky to provide an
impressively precise estimate of peculiar motion associated with the growth of the local super-cluster.
About the same time it was realised that using HI velocity widths for spirals (Tully \& Fisher, 1977; hereafter TF) 
gave distances with $\simeq$ 20\% fractional distance uncertainty.  Central
velocity dispersions for ellipticals (Faber \& Jackson, 1976) provided somewhat less precise
distances but the discovery of the  `fundamental plane' in surface
brightness, velocity dispersion and magnitude (Djorgovski and Davis, 1987; Dressler et al 1987)
allowed the use of the $D_n-\sigma$ relation which uses velocity dispersion
and surface brightness to predict the proper size of the galaxy and which provides distance
precision similar to the TF relation.  
Further peculiar motion data have been obtained from SN1a (e.g.\ Turnbull et al. 2012),
which have much higher precision of $\sim 8$\% fractional distance error per object,
but which are, as yet, a sparser sample of the velocity field than TF or FP samples.
Other methods include using the tip of the red giant branch (TRGB) as a standard candle 
(Lee, Freedman \& Madore, 1993) and surface-brightness fluctuations for early type galaxies which were pioneered by Tonry et al.\ (2000) and
show generally good agreement with FP distances (Blakeslee et al.\ 2002).
All of these techniques are, one way or another, directly measuring the perturbation to the luminosity distance
caused by peculiar motions.

Major reviews of the status of peculiar velocity studies in the mid '90s were provided by Dekel (1994) 
and by Strauss and Willick (1995) and a good snapshot of the state of the subject at the turn of the century is given by
Hudson (1999) and by Courteau and Dekel (2001),
highlighting the results presented at the famous conference in Victoria, B.C. in 1999.
After something of a lull, the subject is now undergoing
a revival, with major advances in the data available, including the `CosmicFlows-2' compilation of
$\sim$8,000 mostly TF distances (Tully et al 2013;
incorporating the SFI++ TF catalog of Springob et al.\ 2007 as well as SNia distances)
and with significant results coming from the 6dFGS survey with 
an impressive $\sim$10,000 Fundamental Plane distances (Springob et al 2013).
In addition, there has been something of a revival in using the individually less precise
galaxy flux densities as distance estimators (Nusser, Branchini and Davis 2011; 
Branchini, Davis and Nusser 2012) exploiting the larger sample size and excellent photometric
accuracy of the 2MASS redshift survey (Huchra, 2012).  Nusser, Branchini and Feix (2013)
have discussed how this might be extended to massive photometric
surveys ($\simeq 15,000$ square degrees with 10-$\sigma$ multi-passband optical through near-IR photometry for
galaxies to $m_r \sim 24.5$) that will be provided by the Euclid satellite along with
the anticipated ground-based support photometry (Laureijs et al.\ 2011).

\subsection{Other observable effects of peculiar motions}

There are other potential probes of velocity that are related to the above but are not exactly
equivalent to measurement of the luminosity distance. 

\subsubsection{The Kinematic Sunyaev Zel'dovich effect}
\label{sec:review-ksz}

One is the kinematic SZ effect on the cosmic microwave background (CMB)
which has been attempted with clusters of galaxies with WMAP by
Kashlinsky et al.\ 2010 and Kashlinsky, Atrio-Barandela, \& Ebeling 2011
and with Planck (Planck Collaboration et al.\ 2014)
and with galaxies (Lavaux, Afshordi, \& Hudson, 2013).  This method has
the distinguishing characteristic that the observable is largely independent
of distance -- unlike standard candle techniques that have poor sensitivity
at large distance -- and should in principle provide the most
powerful constraints on peculiar velocities at large distances.

This method is also quite different from the other probes considered here in that
it provides a remote measurement of the CMB dipole as would have been
seen by observers in the clusters, providing essentially 
unique constraints on departures from homogeneity -- as opposed to more easily measured
departures from isotropy -- of our Universe.

\subsubsection{Dipoles of isotropic backgrounds and source counts}
\label{sec:review-dipoles}

Another technique is to exploit the dipole induced in otherwise
isotropic radiation backgrounds or distant source counts 
by our motion. This was originally applied to cosmic rays
by Compton \& Getting (1935) but is best known, and most accurately measured, from the dipole of the
cosmic microwave background (e.g.\ Hinshaw et al.\ 2009) which indicates a solar motion of 369 km/s.  
The dipole moment of the X-ray background has been measured by Plionis and Georgantopoulos (1999) at 1.5keV and
by Scharf et al.\ (2000) in the 2-10keV energy range, both of whom find results broadly compatible with
the CMB motion.  Lahav, Piran and Treyer (1997) have
shown that in these bands the effects of clustering of sources and that induced by our motion are 
expected to be comparable.

A closely analogous effect is the dipole of counts of discrete sources.
Ellis \& Baldwin (1984) showed that the perturbation to the flux density of
a distant source with spectral index $\alpha$ caused by our motion is $\delta S / S = (1 + \alpha) \times (\bn \cdot \bv /c)$
with $\bn$ the direction to the source.
So, for unresolved sources with $N(>S) \sim S^{-x}$, and allowing
for aberration, the amplitude of the dipole of the
counts is $D = (2 + (1 + \alpha)x)\times (v/c)$. 

Baleisis et al.\ (1998) have attempted to measure this with the Green Bank (87GB) (Gregory \& Condon 1991) and 
Parkes - MIT - NRAO (PMN) catalogues (Griffith \& Wright 1993), but with limited success; they attribute the very
large apparent signal to calibration errors.
The dipole has been determined in the NVSS survey (Condon et al.\ 1998) by Blake \& Wall (2002) 
and more recently by Singal (2011) and Rubart \& Schwarz (2013), the last including
sources from the Westerbork WENSS survey (Rengelink et al.\ 1997). These studies have found generally quite good
agreement of the dipole direction with that of the CMB dipole, but, puzzlingly, finding something 
like 2-4 times the expected amplitude. Gibelyou \& Huterer (2012) have measured dipoles of various
galaxy redshift surveys and also gamma-ray bursts as well as the NVSS radio sources; they
find only the latter to be discrepant with theoretical expectations and suggest the NVSS
dipole may be corrupted by systematic errors.

Mertens et al.\ 2013 have suggested that the Milky Way motion would induce a dipole signal in the magnification
measured in large-area weak lensing surveys such as LSST or DES.  But the quantity that
they consider is the perturbation to the angular size of objects at a constant distance,
which is not directly comparable to what is actually observed. 

The kinematic SZ and source count dipole effects are physically 
distinct from methods that
measure the perturbation to the luminosity distance using flux densities of sources --
or features in the distribution of luminosities like the `knee' in the galaxy luminosity function -- that
can be considered to be `standard candles'. 
These different methods are nonetheless supposed to be measuring the same physical quantity.

\subsubsection{Distortion of galaxy clustering in redshift space}
\label{sec:review-zspacedistortion}

Another manifestation of the perturbation to the distance is the distortion
of the appearance of cosmological structure in redshift space (Kaiser 1987).  
There are two aspects
to this.  The most well known, and most highly exploited, is the distortion of the
correlation function; a combination of smearing along the line of sight from
small scale motions and a coherent squashing associated with large scale motions
(see e.g.\ Hamilton 1997; Percival et al.\ 2011 for reviews).
A separate effect, less widely considered, but more closely related to the
subject here, is the distortion of the density contrast $\delta(\br)$ inferred
from the density of galaxies as this is estimated by dividing the counts
in a region by the selection function $\phi(r)$; this is a function of the
distance but is usually evaluated in redshift space, resulting is a 
biased picture of local structure.
Sometimes referred to as the `Rocket-effect', this effect, like peculiar velocity
measurements, is most important at low redshifts.

In one sense one can consider this to be another probe of peculiar motions, 
but the difficulty is that the velocity induced density perturbation is superposed on
the real-space clustering pattern, and decoupling these is not easy.
The distance measurement methods as described above,
in contrast, provide distances, and hence peculiar velocities, that are not biased or otherwise affected by 
perturbations to the galaxy density.  Such methods give a clean determination of the
velocity, though subject to the assumption that the the luminosity function of the
sources being used is universal (i.e.\ unperturbed by local density or other
environmental influences).  This decoupling of density and velocity is achieved at the cost of
requiring that the estimator, in the case of galaxy-based measurements, may only use information contained in
the {\em shape\/} of the flux density (or angular size) distribution of those objects
that survive the selection criteria; they may not use the actual number of detected
objects.\footnote{There is an analogous situation in lensing at higher redshift where one can measure
the convergence, or magnification, through the perturbation to the
counts of background galaxies or quasars caused by intervening structures 
(see e.g.\ Scranton et al.\ 2005 and references therein for the quasar
studies and Broadhurst, Taylor and Peacock (1995) for a discussion of magnification
of galaxies and Hildebrandt, et al. 2011 for a recent application).
But the noise in this case is enhanced by the clustering of the background
galaxies.  Avoiding this clustering noise is the motivation for the 
convergence estimators of Schmidt et al.\ (2012) and Huff \& Graves (2014)
that are designed to be unbiased by fluctuations in the galaxy number density.
This is harder -- it requires redshift information for the sources and is demanding
in terms of photometric calibration accuracy -- but results, in principle, in
a shot-noise limited measurement of the convergence.}
In the local universe the effect of velocities on the galaxy density field
is not so much a useful probe of structure as an impediment to
attempts to determine the local peculiar gravity field for comparison
with the velocity field with the aim of determining the growth rate of
structure (Nusser, Davis \& Branchini 2014).

\subsection{Perturbation to the luminosity distance from gravitational lensing}
\label{sec:review-gravlensing}

The fractional perturbation to the luminosity distance from peculiar motions
falls rapidly with redshift, but at larger redshift another effect
comes into play which is the modulation of sizes of objects
by gravitational lensing.  This was first considered theoretically
by Zel'dovich (1964) and by Gunn (1967a; citing unpublished work of Richard Feynman
presented at a Caltech colloquium in 1964).  
The essential question is: given a space-time with inhomogeneity and metric fluctuations,
how is the angular size of distant objects modified by the structures along the
line of sight.  This can be obtained by propagating a 
narrow bundle of rays back from the observer to a distant source plane at some specified
distance, which provides the 2x2 matrix giving the mapping from 
source plane positions to sky-plane angles whose determinant is the amplification (the off-diagonal and asymmetric parts of the
diagonal components being the shear and the mean of the diagonal terms being conventionally written as
$1 + \kappa$ with $\kappa$ being the convergence).
At linear order this may be calculated in the Born approximation; the first order relative
deviation of a pair of null rays is calculated using tides evaluated along the unperturbed paths.

The above papers focused on the fluctuations in the convergence caused by structures along the line of sight.
Also of interest is the mean amplification which
will be non-zero if the paths to observed objects tend to avoid foreground objects.  This was
considered by Gunn (1967b) and by Dyer and Roeder (1972, 1974).  

The lensing effect is a cumulative one; the amplification of a background object
depends on an integral of the transverse tidal field along the line of sight.  Consequently the 
convergence changes quite slowly with distance.  Liouville's theorem tells us that
the surface brightness depends only on the redshift, so the fractional perturbation to the
luminosity is, at linear order, just the fractional change in the solid angle, and therefore the
luminosity distance perturbation is related to the convergence by
$\delta d / d = - \kappa(z)$. 

\subsection{More recent theoretical developments}
\label{sec:review-theory}

The early theoretical papers cited above considered universes containing point masses or used `swiss-cheese' models.
A significant advance in understanding of the magnitude-redshift relation in more realistic
perturbed FLRW models with metric fluctuations and motions given according to the prediction for
the spectrum of fluctuations in $\Lambda$CDM or similar models
was provided by Sasaki (1987) and elucidated
by Futamase and Sasaki (1989) and by Sugiura, Sugiyama and Sasaki (1999; hereafter S$^3$99).
These papers provided a unified treatment valid at both low and high redshift and revealed
clearly how the lensing and peculiar velocity effects come to dominate at high and low redshifts
respectively (they also included other, typically subdominant, effects such as the
integrated Sachs-Wolfe (1967) effect).

Ignoring the effects of the peculiar gravitational potential, one finds from S$^3$99's equation 3.15 
that the effect of peculiar velocities of sources `S' and the observer `O' induce a 
fractional change in luminosity distance
(which is to say minus half the fractional change in flux density at constant observed redshift)
\begin{equation}
\left.\frac{\delta d}{d}\right|_z  = -\frac{a}{a'\chi}(\bv_\S \cdot \bn - \bv_\O \cdot \bn) +  \bv_\S \cdot \bn .
\label{eq:DeltadL}
\end{equation}
Here $a$ is the scale factor, $\chi = \int dz / H$ is conformal distance,
prime denotes derivative with respect to conformal time and $\bn$ is a unit vector from the observer along the line
of sight back to the source.  Velocities here and below are in units such that the speed of light $c=1$.

At low redshift, $a / a' \chi = 1/(a H \chi) \simeq 1/z$ 
so the first term in (\ref{eq:DeltadL}) is dominant and is simply the
peculiar displacement $\delta d = - \delta v / H$.  The last term can be safely 
ignored for all of the observations described above, but
becomes significant at finite redshift.  
Interestingly, and indeed rather puzzlingly, this term depends only on the
motion of the source, unlike the leading order low-redshift effect that depends on
the relative peculiar motion of the source with respect to that of the observer.
Thus (\ref{eq:DeltadL}) might appear to say that in a situation where the
observer and the sources share the same common motion, perhaps because they have been
accelerated by some very distant mass excess, there would be an observable dipole
perturbation to the luminosity distance at constant $z$ from which we can infer our motion
relative to the `cosmic frame'.  But that would violate the equivalence principle (EP)
which dictates that for sources and observer in free fall the only locally observable effect
of distant masses is the tidal distortion, which would show up as a quadrupole.  The velocity of the observed
region as a whole with respect to the state of motion of more distant
matter cannot be determined by local measurements.  Yet for $\bv_\S = \bv_\O$ 
equation (\ref{eq:DeltadL}) seems to provide an absolute cosmic speedometer. 

Subsequently, many papers have repeated this kind of analysis.
Pyne \& Birkinshaw (2004) also computed $\delta d / d$ in perturbed FLRW models.
They focussed mostly on gravitational lensing effects at high $z$ but the velocity dependent
terms in e.g.\ their equations 32 \& 23 are equivalent to (\ref{eq:DeltadL}).  
Their analysis makes clear that (\ref{eq:DeltadL}) does not rely
on the assumption that the velocities are purely growing mode
linear perturbations.

Hui \& Greene (2006) performed a similar perturbation analysis, and they
also showed how (\ref{eq:DeltadL}) can be understood heuristically
in the case of an unperturbed FRW model in which sources and the observer
have peculiar velocities (i.e.\ what we are calling `unsupported' motions).  They argue that on all relevant scales
equation (\ref{eq:DeltadL}), augmented by the usual gravitational lensing effect,
provides an adequate description.
Hui \& Greene (2006), Cooray \& Caldwell (2006) and Davis et al.\ (2011)
have all used (\ref{eq:DeltadL}) to calculate the impact of peculiar
velocities on the precision of SN1a cosmology; these papers calculate the extra
contribution to the error covariance matrix over and above the diagonal 
contribution from uncorrelated measurement errors and the corresponding
increase in the uncertainty in the derived dark energy properties. 
 
Hui \& Greene make the important point that even small velocities
can have a big impact because the errors they induce are strongly spatially
correlated.  
Since the primary goal of SN1a is to determine the dark energy `equation of state', 
what is relevant is the monopole of the distance perturbation.  
For an isotropic uniformly sampled full sky survey the observer motion would be irrelevant as it
generates only a dipole.  But for realistic current or near term future
surveys the sky coverage is strongly non-uniform and so there
is coupling of the dipole -- and other low order moments -- into the monopole.
So for real surveys, the motion of the observer -- which couples to all of the
errors at some level -- is the most critical. Yet
all of the papers cited above only compute the extra covariance arising 
from source motions. 
The solution, as suggested by Hui and Greene, is to calculate the correlated motions of SN hosts, 
conditional on the motion of the Local Group motion with respect to the CMB.
They find that this makes little difference for most current SN1a surveys, but may be important 
for measurements that use lots of low-$z$ SN to pin down the current expansion rate more accurately.

Bonvin, Durrer and Gasparini (2006) also calculated $\delta d / d$
in linearly perturbed FRW models, again reproducing the
dependence on velocities of S$^3$99 above (see their equation 59,
and noting that in their notation $(\eta_O - \eta_S) {\cal H}_S$ is the 
same as $a' \chi / a$ in (\ref{eq:DeltadL})).  They suggest that the
power spectrum of $\delta d / d$ provides a new observational tool which can
be used to determine cosmological parameters.  Second order perturbations
to the distance have been considered by Umeh, Clarkson \& Maartens (2012, 2014)
and by Marozzi (2015).

Bonvin (2008) calculated the effect of
peculiar velocities on weak gravitational lensing convergence in perturbed FRW models.
The version of the paper in the journal has an expression for the velocity induced lensing convergence 
that, unlike (\ref{eq:DeltadL}), appears to depend only on the relative
source and observer velocities, but this has been emended in a revised version
on the arXiv.

Following this, Boleko et al.\ 2013 have noted that the standard
lensing convergence effect (that produced by the transverse deflection of rays)
``is swamped at low redshifts by a relativistic Doppler term that is typically neglected''.
They have dubbed this `anti-lensing'.  
The same group (Bacon et al.\ 2014) have followed this with a more detailed
analysis estimating how well this should be measurable with deeper future redshift surveys
and have coined the phrase `Doppler-lensing' to describe the effect
of peculiar velocities.

There is clearly a bit of a disconnect between two disparate communities here. 
The cosmological perturbation theorists are evidently unaware that the 
probes of cosmology that they are proposing have a long and productive history
of practical application going back about 4 decades as described in \S\ref{sec:review-standard-candles} above.
The use of power-spectrum or auto-correlation analysis was also
well developed early on (Gorski, 1988; Kaiser, 1988; Gorski et al.\ 1989; Groth, Juszkiewicz \& Ostriker 1989).
Somehow the equivalence of the physical quantities being discussed 
by the different communities here has been lost somewhere along the line.  The confusion may have
been exacerbated by the description of the velocity induced
effects as `weak lensing convergence'.  There is no lensing here, at least
in the commonly used sense of deflection of light rays by cosmological structure (there
is a deflection of rays caused by aberration associated with motion
of the observer -- but that is largely ignored in the papers above), and
therefore no convergence of rays; all that is happening is that
velocities perturb the relation between observed redshift and angular diameter distance.

As already mentioned, the asymptotic $z \rightarrow 0$  behaviour of equation (\ref{eq:DeltadL})
can be easily understood without recourse to relativity or cosmological perturbation theory; if a source has
a peculiar velocity (relative to any peculiar velocity of the observer) then
it has a peculiar displacement, in redshift space, that is simply $\delta r = - \delta v / H$.
For most of the measurements described above, this simple notion provides a reasonable model
for interpreting the observations.
What is interesting about the
formalism of Sasaki and colleagues (and the more recent theoretical
papers cited above) is not just that it provides a nice
unified treatment of peculiar motions and gravitational lensing, but that
it should be valid to describe the effect of peculiar velocities at all redshifts, not just in the limit $z \ll 1$.
Future redshift surveys will probe to larger redshifts and at even at 
$z \sim 0.1$ the next order correction to the dominant $\sim v / cz$ term --
a contribution to $\delta d / d \sim v/c$ which is independent of redshift --
is a 10\% effect.  Similarly, it would seem prudent to get this
right when computing the impact of flows on scales of a few hundred
Mpc on SN1a cosmology, given how much is riding on the result.  But, as we have noted,
the formalism used exhibits seemingly unphysical behaviour.

In the following section we show that in the case of `unsupported' motions -- that is to say
motions of observers or sources considered as massless test particles
in an unperturbed cosmology -- the resolution of 
the puzzle of the apparent violation of the EP
is that the velocities are decaying, so a spatially uniform flow at a given instant of cosmic time
will appear non-uniform on the past light cone of the observer, and it is
light-cone variables that appear in the formula (\ref{eq:DeltadL}).  The finite
redshift term can be thought of as the correction so that the total effect is
determined by the relative velocities at constant cosmic time.
Allowing for this restores the symmetry between source and observer motions.
For finite matter density and
velocities associated with growth of structure the situation is a
little more complicated, and the resolution is that the EP violating
dipole from the velocity field alone is cancelled in a consistent treatment
by the gravitational redshift effect.
In the process we obtain a somewhat
improved expression for the perturbation to the luminosity distance
that properly accounts for the motion of the observer.
This can be used to better predict the effect of
structure on the precision of the SN1a cosmological parameter estimation.

We then turn attention to the prospects for measuring peculiar velocities from the perturbation
to the redshift as a function of flux density, as proposed by Nusser, Branchini \& Feix (2013).
We show that this has the expected asymptotic behaviour at very low redshift 
that $\delta z / (1 + z) \simeq \bn \cdot (\bv_\S - \bv_\O)$ but that, as
with $\delta d / d$, there are finite-redshift modifications; here caused in part
by relativistic beaming effect modifying the surface brightness.
But the challenge
with this method is that the proposed measurable is swamped for most scales of interest by fluctuations
in the mean redshift caused by large-scale structure.

In the Discussion section (\S\ref{sec:discussion}) we consider the extension of peculiar velocity studies
to larger scales with future redshift surveys.

In Appendix \ref{sec:appendix} we present a linearised analysis of the bias in the velocity field
measurement caused by small scale motions or measurement errors 
and we show that this accounts nicely for the damping found empirically
in the numerical studies of Koda et al.\ 2013.

\section{Perturbation to flux densities at constant redshift}
\label{sec:deltamofz}

First let's recall the origin of (\ref{eq:DeltadL}).
Hui and Greene (2006) have given a very nice heuristic derivation
of this.  They compute the effect of source and observer
velocities in a homogeneous FRW model; i.e.\ they consider the
effect of what we are calling `unsupported' motions.  
They calculate the perturbation
to the object flux density and to the redshift at fixed distance, and
then combine these to obtain the perturbation to $d$ at 
fixed observed redshift.  This is legitimate, but is somewhat roundabout as the
perturbation to the flux densities involves consideration of relativistic
beaming in which the redshift perturbation modulates the surface
brightness of the sources (see e.g.\ the discussion
in Davis et al.\ 2011). A more direct approach is simply to compute the
perturbation to the angular size and redshift to obtain
the `convergence' or equivalently minus $\delta d / d$.

Consider, for simplicity, a spatially flat background cosmology with metric
$ds^2 = a^2(\eta) (-d \eta^2 + d \chi^2 + \chi^2 d \Omega^2)$ where
the scale factor is dimensionless and is equal to unity at present: $a(\eta_0) = 1$
and with an observer at the origin of the spatial coordinate system $\chi = 0$.

In the absence of source or observer motions the 
redshift of a source at conformal distance $\chi$ is $z = \zbar(\chi) = \int d\chi H(\eta_0 - \chi)$ and
the distance is a function of redshift: $\chi = \chibar(z) = \int dz / H$.
The angular diameter of a small spherical standard source of physical diameter $l$ at
this distance is $\theta = \thetabar(\chi) = l / a \chi = (1 + \zbar(\chi)) l / \chi$,
which varies with redshift as
\begin{equation}
\frac{d \thetabar}{dz} = \frac{l}{\chi} \left( 1 - \frac{a}{a'\chi} \right).
\end{equation}
As is well known, this vanishes (at about $z \simeq 1.6$ in the standard $\Lambda$CDM model)
where $\thetabar(z)$ is stationary owing to the focussing of rays by the smoothly
distributed matter.

An observer moving with peculiar velocity $\bv_\O$ relative to a local fundamental observer would
perceive a standard source at distance $\chi$ to have angular size
\begin{equation}
\theta = \thetabar(\chi) (1 - \bn \cdot \bv_\O + \ldots)
\label{eq:theta}
\end{equation}
because of special relativistic aberration.
This is easily shown by applying a Lorentz boost and linearising in the assumed small velocity $\bv_\O$.
There will also be a perturbation to the redshift
\begin{equation}
z = \zbar(\chi) + \delta z 
\end{equation}
where, in the present circumstances, $\delta z / (1 + z)$ is the fractional perturbation to the
wavelength caused by the 1st order Doppler effect of peculiar velocities of the source and observer.

The quantity of interest here is how the perturbed angular size (\ref{eq:theta}) at this
perturbed redshift differs from the angular size for a source at redshift $z$ in the
fictitious case that there is no source or observer motion.  Working to 1st order precision this is
\begin{equation}
\theta_0 = \thetabar(\chibar(z)) = \thetabar(\chibar(\zbar(\chi) + \delta z))
= \thetabar(\chi) + (d \thetabar / dz) \delta z + \ldots
\label{eq:theta0}
\end{equation}

So, subtracting (\ref{eq:theta0}) from (\ref{eq:theta}), and dividing by $\theta$ gives
the first-order fractional perturbation to the angle at this redshift $z$ from peculiar motions as
\begin{equation}
\left.\frac{\delta \theta}{\theta}\right|_z = 
- \frac{d \theta}{dz} \frac{\delta z}{\theta} - \bn \cdot \bv_\O  =
\left(\frac{a}{a'\chi} - 1 \right) \frac{\delta z}{1 + z} - \bn \cdot \bv_\O .
\label{eq:deltatheta0}
\end{equation}

This is (minus) the perturbation to the angular diameter distance, but as this is at
constant redshift this is the same as minus the perturbation to the luminosity distance,
as the surface brightness depends only on the redshift. 
In the situation described here of unsupported peculiar motions in a homogeneous
background cosmology the perturbation to the redshift is 
$\delta z / (1 + z) = \bn \cdot (\bv_\S - \bv_\O)$ which in (\ref{eq:deltatheta0}) is equivalent to (\ref{eq:DeltadL}).

More generally, the fractional wavelength perturbation $\delta z / (1 + z)$ in (\ref{eq:deltatheta0})
will also include contributions from the perturbation to the potential; the peculiar gravity.
But note that for perturbations on scales much less than the Hubble scale, which
is what is relevant here as the perturbations to the velocity on the horizon scale
is on the order of the primary CMB anisotropy or about $10^{-5}$, which is only a few km/s, 
the perturbation to the redshift is well described by Newtonian theory, so 
the only significant `relativistic' effect here is the aberration caused by the observer motion.
And it is this aberration -- or `beaming' -- that is the source of the asymmetry between
observer and source velocities in (\ref{eq:DeltadL}).

Returning to the effect of unsupported peculiar motions, equation (\ref{eq:deltatheta0}) can be written as
\begin{equation}
\left.\frac{\delta \theta}{\theta}\right|_z = \frac{a}{a'\chi} \left[ \left( 1 - \frac{a' \chi}{a} \right) \bn \cdot \bv_\S - \bn \cdot \bv_\O \right] .
\label{eq:deltatheta1}
\end{equation}
This might appear, at face value, to have a zeroth order (in redshift) non-zero value in the case that the
observer and source share the same motion.
But that is illusory.  The velocities appearing in (\ref{eq:deltatheta1}) are at different times,
and in the situation described here the velocities are decaying as (e.g.\ Peebles 1980)
$\bv \propto 1/ a = 1 + z$, so $\bv_\S(\eta_\S) = \bv_\S(\eta_\O) (1 + z)$, and
\begin{equation}
\left.\frac{\delta \theta}{\theta}\right|_z = \frac{a}{a'\chi} \left[ \left( 1 - \frac{a' \chi}{a} \right) (1 + z) \bn \cdot \bv_\S - \bn \cdot \bv_\O \right]
\label{eq:deltatheta2}
\end{equation}
where the velocities are both at the time of observation $\eta_\O$.  But $ 1 - a' \chi / a = 1 - z + {\cal O}(z^2)$ and 
consequently at low redshift
the perturbation to the angular (or luminosity) distance depends only on the relative motion of the source and
observer (with corrections that are quadratic in the redshift):
\begin{equation}
\left.\frac{\delta \theta}{\theta}\right|_z \simeq \frac{a}{a'\chi} \bn \cdot (\bv_\S - \bv_\O)_{\eta_\O}
\label{eq:deltatheta3}
\end{equation}
where, by `$\simeq$', we mean equal up to fractional corrections of order $z^2$.
Thus the apparent EP violation in (\ref{eq:DeltadL}) appears because, in that equation, the 
source and observer velocities are not evaluated at the same time.  When they are,
then to lowest order the EP violation disappears and the symmetry between
observer and source motions is restored.

We are more interested in the case that the motions are associated with the growth of
structure, in which case the velocities do not simply decay as $\bv \propto 1 + z$, rather
the equation of motion for the peculiar velocity is
\begin{equation}
d\bv/dt = - H \bv + \bg
\end{equation}
where $\bg$ is the gravity sourced in Poisson's equation by the matter density perturbation $\delta \rho$ (Peebles 1980).
Consequently $\bv_\S(\eta_\S) \simeq (1 + z) \bv_\S(\eta_\O) - \bg_\S \Delta t$ which, in (\ref{eq:deltatheta1})
would seem to give a zeroth order (in redshift) contribution to $\delta \theta / \theta$ of 
$- \bn \cdot \bg \Delta t / cz \simeq - \bn \cdot \bg / H$.  This is a dipole which, if observable, would
allow one to determine the acceleration from distance structures from local observations,
again in violation of the EP. But if one is going to 
incorporate the peculiar gravity then, to be consistent, one should use, following S$^3$99, for
$\delta z / (1 + z)$ in (\ref{eq:deltatheta1}) not just the Doppler shift, but one should also include the
gravitational redshift $\psi_\S - \psi_\O \simeq \int d{\bchi} \cdot \bg(\bchi)$ which, when multiplied by the large pre-factor $\sim 1 / z$ gives
an effect of the same order for structures on the scale of the source-observer separation.
We then have
\begin{equation}
\left.\frac{\delta \theta}{\theta}\right|_z \simeq \frac{a}{a'\chi} \bn \cdot (\bv_\S - \bv_\O)_{\eta_\O}
- \frac{\bn \cdot \bg_\S}{H} + \frac{1}{z} \int\limits_{\bchi_\O}^{\bchi_\S} d{\bchi} \cdot \bg(\bchi) .
\label{eq:deltatheta4}
\end{equation}
For the case of a distant attractor the gravity $\bg$ will be nearly
spatially constant within the observed region and since $\bn = (\bchi_\S - \bchi_\O) / |\bchi_\S - \bchi_\O|$
the last two terms cancel and the gravity from distant matter is therefore unobservable from
local measurements, again in accord with the EP. 

In addition to being blind to super-survey size sources of gravity,
expression (\ref{eq:deltatheta4}) has the desirable characteristic is that
it is independent of the choice of `background' that we consider the actual universe to be a perturbation about. 
For example, were one to consider the observable region
to be a perturbation to a background with the ``wrong'' density and therefore the wrong value for the Hubble
parameter, say, this would change the peculiar velocities.  But the corresponding spatially constant density perturbation
would produce a spatially constant tidal field which would compensate so that the
physical observable $\delta d / d$ is invariant.

Finally, in linear theory the gravity and velocity are proportional to one another, with $\bv = (2 f(\Omega_m) / 3 \Omega_m) \times \bg / H$,
where the perturbation growth rate factor $f(\Omega_m) \equiv d \ln \delta / d \ln a \simeq \Omega_m^{0.55}$,
and we can express the perturbation of the luminosity distance entirely in terms of velocities as
\begin{equation}
\begin{split}
\left.\frac{\delta d}{d}\right|_z  & \simeq - \frac{a}{a'\chi} \bn \cdot (\bv_\S - \bv_\O) \\
& + \frac{3 \Omega_m}{2 f(\Omega_m)} \left( \bn \cdot \bv_\S - \frac{H}{z} \int\limits_{\bchi_\O}^{\bchi_\S} d{\bchi} \cdot \bv(\bchi) \right)
\end{split}
\label{eq:deltadd}
\end{equation}
where the velocities are all to be understood to be at the time of observation.

This formula (\ref{eq:deltadd}) provides, at least in the context of linear
theory, a physically consistent way to allow for the effect of peculiar motions on
supernova cosmology.  Given an assumed power spectrum for the density, and hence
velocity, perturbations, it can be used as in the studies referenced above
to compute the extra contribution to the covariance function.  Also, given some
estimate of the actual structure in the nearby universe, it can be used to correct
the distances of SN for the associated motion, with the additional covariance
from larger scale motions being calculated from the conditional probability
distribution for the residual motions.

\section{Perturbation to the redshift at constant flux density}
\label{sec:deltazofm}

Nusser, Branchini and Feix (2013) have suggested that
useful measurements of large scale motions could be obtained
using the photometric redshifts from Euclid (Laureijs et al.\ 2011)
which  will be obtained from a combination of ground-based
surveys (providing at least three of the $g$, $r$, $i$ and $z$ bands)
and the Euclid photometry in the broad VIS band and in $Y$, $J$ \& $H$.
This is expected to provide $\sim 10^9$ photometric redshifts with better than $\sim$ 5\% precision
in $1+z$ to mean redshift $\zbar \simeq 1$ and covering 15,000 square degrees.
The ground based surveys will perform photometry with $\sim 10$-sigma detections limits 
deeper than 24th magnitude and may by themselves provide
useful photometric redshifts in advance of Euclid.  Indeed some of these data in the Southern
sky are already being collected by the DES\footnote{http://www.darkenergysurvey.org/} dark
energy survey which is planned to cover $\sim$ 5,000 square degrees.

The Nusser, Branchini and Feix (2013) proposal differs from conventional peculiar velocity  measurements in that rather
than the considering the perturbation to the distance, and therefore the flux densities of objects, at a given redshift
they are proposing to measure the perturbation to the redshift, as a function of direction,
relative to the mean redshift for sources of a given flux density. 
To this end they define a redshift perturbation per galaxy
\begin{equation}
\Theta_i = \frac{z_i - z_{\rm cos}(m_i)}{1 + z_i},
\label{eq:NBF}
\end{equation}
where $z_i$ is the redshift of the $i$th galaxy and $z_{\rm cos}(m)$ is the mean redshift for galaxies of
apparent magnitude $m$.  This, they argue is dominated, for sub-horizon scale structures, 
by the Doppler shift, so that the average of $\Theta_i$ over galaxies lying in some solid angle on the sky
provides an estimate of an appropriately weighted average of $\bn \cdot (\bv_\S - \bv_\O)$ along the
line of sight.
What they propose is to use the angular power spectrum of this quantity to constrain the power spectrum of matter fluctuations.

The motivation for this is that they claim that this would provide measurements of large-scale motions with
a precision corresponding to a fractional distance error of approximately 30\% 
per galaxy from photometry and relatively low resolution spectroscopy.  This is not very much worse than the precision obtained
from the considerably more expensive TF or $D_n-\sigma$ techniques.

However, this technique differs from the normal method in several important respects:
One is that the finite-redshift behaviour is somewhat different from that
for the luminosity distance. 
A second is that by using only the angular variation of the line-of-sight averaged
peculiar velocity this method discards useful information about the variation
of motions along the line of sight (this is significant since it is the angular
variations that are most susceptible to systematic errors in the photometric
zero-point).
But the most important distinction is that the estimator will tend to be
heavily influenced by large-scale structure in the galaxy spatial distribution.
In fact, as we will show, the signal from peculiar motions will be swamped
by fluctuations from galaxy clustering.

The assumption underlying the idea that an average of (\ref{eq:NBF}) provides an
estimate of the average peculiar velocity is that galaxies in some direction 
selected by flux density will have a distribution of distances $P(d)$ that is unbiased
by peculiar motions.  But that is not the case.  Large-scale motion will
affect $\Sigma$, the surface brightnesses of the galaxies, and therefore also their flux densities at a 
given distance.   For
bolometric observations the effect is the usual $(1 + z)^{-4}$ dimming, so
$\delta \Sigma / \Sigma \simeq - 4 \bn \cdot (\bv_\S - \bv_\O)$.
For filter band-pass limited observations there is also a linear relation, but the
coefficient depends on the spectral energy distribution (SED) of the sources
and on the filter transmission function.  Here, for simplicity, we will consider bolometric observations.

If one considers a population of standard objects of identical proper size $l$ and rest-frame surface
brightness $\Sigma_0$ then, in the absence of motions, the flux density from a source at distance $\chi$ is 
\begin{equation}
F_0(\chi) \sim \Sigma_0 d\Omega / (1 + \zbar(\chi))^4 \sim \Sigma_0 l^2 / [\chi^2(1 + \zbar(\chi))^2],
\end{equation}
where `$\sim$' means ignoring geometric factors of order unity and where we
have used the solid angle $d\Omega \sim l^2 / a^2 \chi^2$ as appropriate for
a spatially flat universe.
In the presence of unsupported motions, the flux density for a standard source at the same distance $\chi$ is
\begin{equation}
F(\chi) = F_0(\chi) \times (1 - 2 \bn \cdot \bv_\O) (1 - 4 \bn \cdot (\bv_\S - \bv_\O))
\end{equation}
with the first factor being the fractional change in solid angle from observer motion
induced aberration and the second being the fractional change in the surface brightness.

In general $F(\chi) \ne F_0(\chi)$, rather requiring $F(\chi)$ to be equal to the 
`background' flux density for some perturbed distance $F_0(\chi + \delta \chi)$ gives
\begin{equation}
\delta \chi = \bn \cdot (2 \bv_\S - \bv_\O) / (1 / \chi + H / (1 + z)).
\end{equation}
Such a displacement in the background cosmology would correspond to a change in redshift $\delta z = H \delta \chi$.
Adding this to the first order Doppler shift $\delta z = \bn \cdot (\bv_\S - \bv_\O) (1 + z)$ at fixed $\chi$
gives the fractional perturbation to the observed wavelength for objects at fixed flux density as
\begin{equation}
\left. \frac{\delta z}{1 + z} \right|_{F_0} = 
\frac{(1 - a' \chi / a) \bn \cdot \bv_\S - \bn \cdot \bv_\O}{( 1 + a' \chi / a)} .
\label{eq:deltaz}
\end{equation} 

For $z \rightarrow 0$ the RHS becomes simply the sum of the source 
and observer Doppler shifts (recall that $a' \chi / a \simeq z$ for low $z$).  But for finite
$z$ there are corrections coming from the relativistic beaming effects.
These reduce the observable effect relative to the naive prediction.

The numerator here is identical in its dependence on velocity to the fractional perturbation
to the angular size of standard objects at constant redshift (\ref{eq:deltatheta1}),
so it can also be shown to be only dependent on relative velocities at constant
proper time (plus corrections of order $z^2$) for the case of unsupported motions and
can be cast into a physically consistent form in the case of motions associated with
growing structure with the inclusion of the Sachs-Wolfe term.

This is an idealised model.  The assumption of standard candles means that observations at some $m$
probe only a thin shell at the corresponding redshift.  Allowing a distribution function 
for the absolute magnitude and surface brightness results in a distribution $N(z | m)$
but with the same general dependence on redshift.  As mentioned, the 
details also depend on galaxy SEDS and filter transmission functions.

So far this technique seems quite promising, particularly if the scatter in redshift
is as small as the $\sim$30\% claimed.  Unfortunately, however, 
the estimator (\ref{eq:NBF}) for motions also differs from conventional luminosity
distance estimators in that it is not unbiased by the density of
galaxies. This has negative consequences.  In a universe with no structure, but with moving sources and observers,
the average of (\ref{eq:NBF}) would provide a probe of motions with the relatively
good precision indicated from the width of $N(z)$.  But in reality,
even in the absence of peculiar velocities, there will be additional fluctuations in the
mean redshift along each line of sight caused by density perturbations.

If one were to compute the average of $\Theta_i$ for all of the galaxies in
a cone of small solid angle $d \Omega$ and in some modest range of magnitudes
$dm$ around $m$ as ${\overline \Theta} = N^{-1} \sum \Theta_i $ then the cosmological signal would be
essentially just the mean of the line of sight source velocity weighted by the
density of galaxies (and the finite redshift terms).  But with large scale
structure there will be an additional contribution
\begin{equation}
{\overline \Theta} = (1 + \zbar(m))^{-1} \int d\Omega dz z^2 n(z) (z - \zbar(m)) / N .
\end{equation}
The integral here will, on average, be zero, but there will be fluctuations
arising from the fluctuations in the  number density of galaxies $n(z) = \nbar(z) (1 + \delta(z))$.
In a simple model of `blobs' with some characteristic scale $\Delta z$ and rms density contrast $\delta$ then
the rms contribution to ${\overline \Theta}$ from structure will be equal to $\zbar(m) \delta / \sqrt{N_s}$
times some factor of order unity, where $N_s \simeq \zbar / \Delta z$ is the number of
random structures along the line of sight. 
The `signal', on the other hand, is on the order ${\overline \Theta} \sim \delta \Delta z / \sqrt{N_s}$.
So for structures much smaller than the depth of the survey -- which for deep surveys means
much less than the Hubble scale -- the signal will be swamped by the noise from 
large-scale structure.

Of course the density of galaxies in redshift space is measurable, so one can try to
correct for this bias.  But this leads one back to an estimator
for the luminosity distance perturbation as a function of redshift where
one has to infer the distance from the shape of the distribution of fluxes of
objects above the flux limit (perhaps sub-divided by distance independent properties),
and the statistical precision will not be the same as that of (\ref{eq:NBF}).

\section{Discussion}
\label{sec:discussion}

\subsection{Summary of major results}

To recapitulate the major results so far, in Section \ref{sec:deltamofz} we showed how the somewhat puzzling dependency of the
finite-redshift corrections to the luminosity distance perturbations
on the absolute motion of the source and observer is removed once 
allowance is made for the time evolution of the velocity field and
for the gravitational redshift.  We note that
Sasaki and colleagues never separated the different contributions, so
this does not cast any doubt on the validity of their results. The
only question is regarding the legitimacy of using the peculiar
velocity sourced effects alone. We obtained
an expression for $\delta d/d$ in terms of the peculiar velocity that is manifestly respectful of
the equivalence principle and which provides a consistent way to 
allow for the motion of the source.  This provides an improved way to estimate
of the impact of motions on SN1a cosmology precision.  But this is 
not a huge effect; for local perturbations on scale of a few hundred Mpc this
is a 10\% correction, and in the standard $\Lambda$CDM model there is not
much effect from larger scales.  Nor is this a very big effect for
peculiar velocity measurements (or `Doppler-lensing'; we hope we have dispelled any
notion that these are different) if one is measuring the effect of
perturbations on scales much smaller than the typical observer-source separation,
but again (\ref{eq:deltadd}) should provide a more accurate result and so should be
used in forecasting the performance of future surveys.

In Section \ref{sec:deltazofm} we considered the perturbation to the redshift 
for sources of a given flux density.  We showed that one needs here to consider the
effect of the motions on the surface brightness which results in
a non-trivial dependence of the effect on the
depth of the survey.  But we also argued that the signal would, except for very large-scale
modes, be swamped by galaxy clustering (as this measure of velocity is not unbiased
with respect to galaxy number density).

\subsection{Prospects for peculiar velocities in larger, future surveys}

We turn now to the prospects for extending peculiar velocity studies to greater
depth, with particular focus on the possibility of exploiting future
photometric surveys, possibly augmented with low resolution spectroscopy to
provide redshifts, but without the more expensive velocity width information
required for TF or $D_n - \sigma$ measurements.

\subsubsection{Scaling of FOM with depth}

Extending peculiar velocity studies to large distances generally has a rather
poor return on telescope time because the distance errors grow linearly with depth $D$.
And surveys without velocity dispersions for early type galaxies or HI velocity widths for spirals have
generally higher distance error per galaxy.
But there is still potentially
useful information.  The number of regions of a given size of interest $L$ grows as $D^3$ so the net
error on the variance of flows on scale $L$ does decrease (but only as $1 / \sqrt{D}$ so the
figure of merit (FOM) scales linearly with depth -- or only as the cube-root
of the number of galaxies surveyed).  The measurement of Tammann, Yahil and Sandage (1979)
using luminosities of galaxies as a distance indicator
gave a measurement of the coherent peculiar motion on the $\sim 10 {\rm Mpc}/h$ scale of the local supercluster with 
$\sim$ 100 km/s precision.  So scaling up to a survey volume with huge numbers
of such structures tells us that
obtaining significant measurements of the variance
of flows on similar scales should be do-able. Plus, of course, in addition to making
more precise estimates of the velocity power spectrum at fixed scale, deeper
surveys are invaluable for measuring bulk motion on the scale of the entire survey volume.
It might be hard to justify
mounting a survey on these grounds, but if they are carried out for
other reasons then there is no question that they should yield information on peculiar velocities.

\subsubsection{Peculiar velocities {\it vs.\/} redshift space distortions}

It is reasonable to ask
whether the information thus gained will be competitive with or complementary to that garnered from
redshift space distortion studies (RSD), for which the FOM scales as the volume.  
At the present, these two techniques are roughly equally statistically powerful for measurement of the growth rate parameter $\beta$, despite there being
many, many more galaxies in the surveys used for RSD ($\sim 800,000$) as compared to the
samples of currently $\sim$ a thousand distance measurements (Hudson \& Turnbull 2012; 
Koda et al.\ 2013).  The reason is that RSD measurements are an auto-correlation
measurement and therefore suffer from cosmic variance whereas velocity based
determinations exploit cross-correlation and are therefore free of cosmic variance.
Future surveys such as TAIPAN (Koda et al.\ 2013) and WALLABY (Duffy et al.\ 2012)  with tens of thousands of distances
will yield significant improvement in peculiar velocity measurements 
(see e.g.\ Koda et al.\ 2013).
But ultimately in the future the balance will shift  in favour of RSD because 
distance-based velocities scale poorly with depth as noted above
(kinematic SZ measurements, however, do not suffer this penalty).

A further argument for measuring peculiar velocities as well as RSD is that both are affected by small
scale motions, but in somewhat different ways.  Modelling RSD involves the
pairwise velocity distribution as a function of projected separation
which is difficult to predict.  It is dangerous to
assume a dispersion that is independent of projected separation 
as in e.g.\ the Landy, Szalay \& Broadhurst (1998) method since it is
expected that, qualitatively speaking, the dispersion at
small projected separation is dominated by the `1-halo' contribution
which is heavily weighted to massive clusters while at larger separation
one is dealing with pairs that are in separate virialised systems where
the appropriate average is dominated by small systems.  One can
try to estimate the pairwise velocity dispersion from group and cluster
catalogs, but the result from e.g.\ the Crook et al. (2007)
group and cluster catalogue from 2MRS is small and seems to be at odds with what
is needed to match the data. This, of course, may be simply because this 
simple modelling does not include the relative motions of groups and clusters.  

These small-scale peculiar motions also cause a bias in measurements of larger-scale
flows.  Peculiar velocity measurements suffer are well known to be susceptible
to biases of various kinds (see e.g.\ Faber et al.\ 1994; Strauss \& Willick 1995)
that are often described as `Malmquist' bias.
These biases depend on how the data are selected and analysed (e.g.\ whether one regresses
magnitude onto line-width or {\it vice versa\/}) and on what quantity
one tries to measure (e.g.\ peculiar velocity as a function of distance
{\it vs.\/} peculiar displacement as a function of redshift; or in Strauss and
Willick's terminology method I {\it vs\/} method II).

Peculiar velocity
measurements may be also be biased by environmental (i.e.\ density) dependence of the luminosities
of galaxies.  But in the context of measurements of large-scale linear structure these
can in principle
be measured by cross-correlating apparent peculiar velocity and density in Fourier space.
The true peculiar velocity is 90 degrees out of phase with the density and
so appears as an imaginary contribution to the cross-spectrum while linear environmental
bias shows up in the real part.  See McDonald 2009 for further discussion.

The most serious `Malmquist'-type  effects can be avoided if one
uses the flux densities, together possibly with auxiliary distance independent quantities such as velocity widths or dispersions, 
to estimate the distance to a collection of galaxies in a localised region of redshift space
(i.e.\ method II). For example, if there is no selection on velocity width, as is often
a reasonable approximation, and if residuals in log velocity width are approximately Gaussian, 
as is again often assumed, then an unbiased estimate of the peculiar
displacement is obtained from the deviation of the mean log velocity width from that
expected if the galaxies are at the distance indicated by their redshift.  This is
known, for historical reasons, as the `inverse' method.
More generally, if the joint distribution of absolute magnitude and velocity width,
or other distance independent attributes,
is known then one can obtain the likelihood of the distance under the assumption
that all of the galaxies in that region lie at the same distance, and thereby obtain
an unbiased maximum likelihood estimate for the distance.

These results rely critically on the assumption that the velocity field is locally `cold', so
the galaxies in question can be assumed to have a common distance.
Motions associated with small and intermediate scale structure
violate this assumption and this
introduces a bias not dissimilar to that which effects RSD.  The effect is easy to understand; if one is measuring the
flow in a region of redshift space on the back of a supercluster where the
density of galaxies will be falling with radius and if there is any dispersion
in distance at a given redshift then in a region of redshift space there will
tend to be more galaxies that have scattered away from us than have scattered towards us from behind.
In the Appendix 1.\ we show that in linear theory, one would expect this to cause the measured velocity in redshift space
to be biased downwards by a factor $1 - k^2 \sigma_v^2 / \beta H^2$.

In principle (though subject to having a viable model
for the biasing of the space distribution of galaxies) this velocity bias,
and hence $\sigma_v$ also,
is measurable.  So peculiar velocity measurements provide independent constraints
on the distribution of relative motions of galaxies that is the
major nuisance factor in the interpretation of RSD.

Another potentially useful difference between these probes is that RSD measures only that component of the large-scale
velocity field that is associated with the large-scale density contrast.  In
conventional models for structure formation that is all there is, but in principle
one could imagine that the actual cause of peculiar motions is not given
by the gravity predicted by the density contrast either because of
some kind of exotic `biasing' (e.g.\ large-scale astrophysical modulation of galaxy
creation) or some exotic source of gravity.  
It would be nice to be able to test this aspect of the prediction of gravitational instability directly.

\subsubsection{Peculiar velocities without velocity widths: the `photometric plane'}

Turning to the precision that might be obtained from future surveys without
velocity-width information, Bacon et al. (2014) made forecasts for `Doppler-lensing' experiments
with DES, Euclid \& SKA.
The most conservative example they
considered was the DES photometric survey with spectroscopic follow-up to 
obtain redshift for those galaxies at $0.1 < z < 0.3$.
They assumed a surface density of 0.7 objects per square arc-minute, and
that distances could be obtained with fractional error per galaxy of
30\%.  These numbers both seem a little optimistic.  

This redshift range and surface density implies a space density of
$\simeq 8\times 10^{-2} ({\rm Mpc}/h)^{-3}$ but to reach that density
requires observations of galaxies about 4 magnitudes fainter than $M_*$,
or about 24th magnitude (i.e.\ about as deep as the 10-$\sigma$ photometric
limits -- though only a sub-set selected by photometric redshift to lie
at this low redshift would need to be targeted).  To obtain spectra this
faint requires $\sim 1$ hour integrations on a 10m telescope.  The surface
density is well matched to e.g.\ DEIMOS on Keck-II but the problem is that
the field of view of this instrument is less than 100 square arc-minutes so to cover the 
an area the size of the DES footprint would
take $\sim 10^5$ hours, which is impractical. 

The 30\% fractional distance error estimate was based on the small scatter
in sizes of objects in the HST COSMOS survey found by Schmidt et al.\ (2012) who
devised a weak lensing convergence estimator based on the size and magnitude
of these objects. 
But scatter in observed size does not simply equate to the
distance precision as there are efficiency factors arising from incompleteness
(Schmidt et al.\ 2012 attempted to estimate these).

If one were simply to use the flux densities of all
galaxies in a flux limited survey taken willy-nilly without any
auxiliary information the distance error is very much higher (and the figure of
merit scales inversely as the square of this).  
An interesting question is whether one can do significantly better
using information such as size (or surface-brightness),
morphology and colour.
In this vein Tammann, Yahil \& Sandage divided their galaxies by morphologically
determined `luminosity classes' and similarly Nusser, Branchini \& Davis (2011)
sub-divided the 2MRS galaxies according to their
classification as spirals or ellipticals.  These two sub-samples have similar shaped 
LFs but $M_*$ differs by about 0.4 magnitudes so combining velocity estimates
obtained from these separately has slightly better performance than taking them
together and ignoring the difference in their LFs.

For elliptical galaxies it may be possible to improve the precision
of photometric distances using other distance independent information.
Kormendy (1977) showed that the size of elliptical galaxies is strongly
correlated with their surface brightness, and applied this to measure
distances to the Virgo cluster.  The small scatter in this `Kormendy relation'
would indicate something like 30\% fractional distance uncertainty.
Adding colour information, de Carvalho \& Djorgovski (1998) claimed
distance precision of 25\% and argued that it might be possible to
get 15\% distance uncertainty with more homogeneous data, which
would be better even than the $D_d - \sigma$ relation.\footnote{Scodeggio et al.\ (1997) found a still smaller scatter
in the Kormendy relation for E and S0 galaxies in clusters, but their
result is not directly comparable and, as they pointed out, does not provide a distance indicator.}

This is all very encouraging, but subsequent studies suggest that
the promised performance from the `photometric fundamental plane' 
may be hard to achieve. 
Graham (2002) showed that for early type galaxies
the Sersic index correlates with velocity dispersion giving another
photometric method for distance estimation.  He fit for log effective radius
as a function of surface brightness and Sersic index. However, he found
fractional rms distance errors of $\simeq$ 38\% (E-galaxies only) or 48\% (E + S0).
This was using HST photometry at relatively low redshift, so that may
be optimistic.  La Barbera et al.\ (2005) performed the same kind
of analysis for a pair of clusters at $z \sim 0.3$ and obtained
a tighter relation with $\simeq 31$\% fractional distance error. 
Huff and Graves (2014) fit for a photometric fundamental plane
for early type galaxies (those with best-fit photo-$z$ template
being passive) in SDSS using surface brightness and concentration
index (SDSS did not measure Sersic index).  They used this
to measure weak lensing convergence but if used as a distance
indicator would have a precision of approximately 40\%.

These studies therefore find relations with larger
scatter -- and therefore poorer distance estimate performance -- 
than some of the earlier studies indicated might be
achievable.  And even these results are probably over-optimistic
as most of them are affected by selection effects.  For example, the
Kormendy relation for ellipticals is that the surface brightness is
a decreasing function of size.  There is no doubt that the absence
of large galaxies with high surface brightness is real,
but the absence of small galaxies of low surface brightness is in
part at least a selection effect.  This selection bias acts to artificially
tighten up the relations, inflating the apparent distance precision.

A more realistic estimate of the precision that may be obtainable
requires analysis of surveys with well defined selection criteria.
Smith, Loveday and Cross (2009) have measured the bivariate luminosity-surface brightness
distribution in the UKIDS LAS.  They show that with a optical $u-r$ colour split both
blue and red subsamples are quite well described by Choloniewski models (Choloniewski, 1985).
The red sub-sample has nearly luminosity independent surface brightness and has
a faint end slope $\alpha$ close to zero and much larger than the value of
$\alpha \simeq -1$ characteristic of the galaxy population as a whole.  The bigger $\alpha$ the
better the precision per galaxy, and $\alpha = 0$ gives  approximately 50\% error
in distance per galaxy, so this kind of precision should be obtainable for the red, primarily early-type, sub-sample, and 
the above studies suggest that including morphological information should improve this.
But the downside is that a large $\alpha$ means that one cannot exploit the large
numbers of sub-$L_*$ galaxies that are found for later types.
For $\alpha = 0$ the space density rapidly converges to $\phi_*$, which for these objects is 
$\simeq 1.1 \times 10^{-2} ({\rm Mpc}/h)^{-3}$
and one reaches a density of $\sim$ half of this by going about half a magnitude below $M_*$,
or to about $m_r \simeq 21$, and this would be a more reasonable goal
for spectroscopic follow-up with the 400 fibre 2df/AAOmega spectrograph on the AAT\footnote{http://www.aao.gov.au/2df/aaomega/aaomega.html} perhaps augmented by 
the 1000 fibre WEAVE spectrograph on the WHT\footnote{http://www.ing.iac.es/weave/} if this goes ahead, but it would still be a massive effort.
If this were done, it would provide then $\sim$ 50\% distance errors (rather than 30\%) for objects within
$z \simeq 0.3$ but these would have a space density of only $\simeq 6 \times 10^{-3} ({\rm Mpc}/h)^{-3}$
or a little over a factor 10 smaller than assumed by Bacon et al.\ 2014.  So overall this results in
a big decrease in the FOM.

To obtain high space density at $z \sim 0.3$ as anticipated by Bacon et al.\ 2014 would require advanced spectroscopes
such as PFS\footnote{http://sumire.ipmu.jp/en/2652} or the Maunakea Spectroscopic Explorer (MSE)\footnote{http://www.cfht.hawaii.edu/en/news/MSE}
that combine very wide field with the reach of a 8-10m scale telescope,
but of course there would be competing demands for these, and the currently proposed surveys
for e.g.\ PFS would not be suitable.  In the nearer term it will be possible to try to do peculiar velocity
measurements with photometric redshifts using DES.  But the challenge there is that with 5\%
errors in $1 + z$ the effect of bias described in the Appendix is expected to be
strong even for very large-scale perturbations, and the signal on smaller scales will
be corrupted.  In addition, there is the concern regarding the impact of catastrophic photometric redshift
errors.  An alternative to DES, though likely delivering data on a slightly longer time-scale, is the J-PAS\footnote{http://j-pas.org/}
survey with 56 narrow band filters that, it is claimed, will give photometric redshifts with
$\simeq 0.3$\% precision in $1 + z$ to $i_{\rm AB} \simeq 22.5$ over 8,000 square degrees.  

In summary, the prospects for extension of peculiar velocity studies to
larger scales using photometric surveys, perhaps augmented with low resolution spectroscopy,
are interesting but will not deliver the kind of $\sim 30$\% fractional distance precision
that has been suggested in some previous studies.  The spatial sampling density
will also be an issue, and it will be interesting to see what information can be gained
from `photometric-plane' studies of the more numerous late-type galaxies.  
And without spectroscopy, the measurements
will be limited to very large spatial scales.  Nonetheless, one can expect to obtain useful
information as a spin-off from these surveys that are being carried out primarily for weak-lensing
and for RSD/BAO measurements. 

But, more conventionally,
we can expect to see good progress at lower redshift (where the signal per volume element is much stronger)
with the TAIPAN spectrograph that is being developed for the UKST Schmidt telescope
and which represents a significant advance on the current 6df instrument and which
will provide FP distances as well as redshifts, and from the ASKAP WALLABY survey.  
These will both extend the range of spatial
scales that peculiar velocities probe and will also improve the signal to noise at lower
redshift by increasing the density of sampling.

\section{Acknowledgements}

We are grateful to the referee for helpful suggestions.  MH acknowledges the support of NSERC.
NK is grateful for the support of Cifar and for the hospitality of the Institute for Advanced Studies at Durham.

\appendix

\section{Bias in large-scale peculiar velocity measurements}
\label{sec:appendix}

One way to analyse cosmic-flow data would be to consider each galaxy
to provide a measure of the peculiar velocity at a position in space
which is the estimated distance. If one lived in a rich cluster then
mapping velocity vs.\ estimated-distance space coordinates would make
considerable sense.  But for large-scale peculiar velocity studies
this would be subject to biases of various kinds.  At cosmological
distances redshift gives a more faithful representation of real space
than estimated distance space (where the errors increase linearly with
distance).  If one considers the offsets of the flux densities or
magnitudes from their expected values as giving an estimate of the
peculiar displacement as a function of redshift space position
(Method II in the terminology of Strauss \& Willick 1995)
then a lot of the bias problems are ameliorated.

But it is not completely free of bias.  If you imagine galaxies as
having random motions, much like molecules in a gas, (or large random
measurement errors, as is the case for photometric redshifts) then, if
you look in a localised region in redshift space (perhaps a thin shell
within a cone of some solid angle) then, if there is, for instance, a
radially outwards increasing density of galaxy density then there will
be more galaxies in the selected region that are scattered in from
greater distances than from the nearer side.  That results in a
positive bias in the estimated luminosity distance at the chosen
redshift that is proportional to the radial gradient of the density of
galaxies. 
This bias is present even in a homogeneous universe at low redshift because the number density of galaxies increases outwards $\propto z^2 dz$. The bias arising from this effect was first considered by  Lynden-Bell (1992) and is explored more recently by us in a separate paper
(Kaiser and Hudson 2015). 

There is a related effect arising due to the presence of density inhomogeneities, where, for example, on the near side of a supercluster the density is rising more quickly than in a homogeneous universe and so there will be more galaxies scattered down from greater distances than scattered up from nearer distances.
This effect 
is opposite in sign to that produced by peculiar
motions, which are proportional to (minus) the radial gradient of the
potential.  
The peculiar displacement is $r = - v_{\rm ran} / H$.  If this has a
probability distribution, independent of position, $P(r)$ then the
mean peculiar displacement of galaxies is $\rbar = \int dr r (1 +
\delta_g(r)) P(r) / \int dr (1 + \delta_g(r)) P(r)$ where
$\delta_g(r)$ is the real-space galaxy density contrast.  If there is
only a large-scale density perturbation then one can locally expand
$\delta_g$ as a Taylor series and, since the mean of the distribution
of $r$ is taken to be zero, we have $\rbar \simeq (d \delta_g / d r)
\langle r^2 \rangle$ corresponding to a peculiar velocity $- H \rbar$.

For a plane wave, $\delta_g(\br) = \delta_g \exp(i \bk \cdot \br)$ the
inferred peculiar velocity is
\begin{equation}
v_{\rm meas} = v_{\rm phys} + v_{\rm bias} = v_{\rm phys} \times (1 -   k^2 \sigma_v^2 / \beta H^2) .
\label{eq:velbias}
\end{equation}
So if the velocity dispersion is known, one can estimate, and correct
for, the bias on any scale of interest (noting, of course, that
wavelength and wave number are related by $\lambda = 2 \pi / k$).

This bias -- or `damping' -- of the measured peculiar velocity will
tend to bias estimates of $\beta$ from comparison of density and
velocity fields.  In the analysis of Willick \& Strauss (1998) the
effect of small-scale motions was included in the model.

It is worth noting that that a similar bias also affects peculiar
velocities measured at a given estimated distance (Method I).  
In this case the
bias is known as ``inhomogeneous Malmquist bias'' (Hudson 1994,
Strauss \& Willick 1995).  It has the opposite sign as the bias
considered above since uncertainties in the distances cause the estimated real-space positions of galaxies to scatter out of overdensities, leading
to an apparent peculiar velocity signal that has the same sign as the
infall signal.  Usually this bias is eliminated by using the galaxy density
field to correct individual distance estimates.  However it is also
possible in principle to make a statistical correction on a
mode-by-mode basis.  In this case, the bias is very similar to the one
described above, but with opposite sign:
\begin{equation}
v_{\rm meas} = v_{\rm phys} + v_{\rm bias} = v_{\rm phys} \times (1 +  k^2 \sigma_d^2 / \beta) .
\label{eq:velbiasrealspace}
\end{equation}
Note that for most realistic surveys $\sigma_{d} \gg \sigma_{v}/H$,
hence this bias is larger than the Method II bias.

Returning to the case of peculiar displacement in redshift space, it
is clear that the model for random motions discussed above is rather
idealised.
It may be a reasonable model for redshift measurement errors, but real
galaxy motions are a complicated mix of virialised velocities within
bound systems and streaming motions on larger scales.  We would argue
nonetheless that this model should be applicable -- provided one is
dealing with large-scale waves in the linear regime -- to describe the
effect of small $\sim {\rm Mpc}/h$ scale motions in virialised groups
and clusters.

Whether one should include the effect of motions on intermediate
scales depends rather on how the data are analysed.  A very large
scale density enhancement (the scale of interest) by itself causes a
smooth deformation of the physical distance shell corresponding to the
chosen constant $z$.  Perturbations on some smaller scale (which we'll
assume here is either linear or quasi-linear so position and redshifts
are single-valued functions of one-another) will further corrugate the
constant-$z$ distance surface.  If one makes an estimate of the very
large scale velocity by averaging the flux densities of the galaxies
then the intermediate scale perturbations will bias the result in the
manner of (\ref{eq:velbias}).  This is because, again in the case that
the large-scale mode has a positive radial density gradient, wherever
the intermediate scale motions result in a positive peculiar
displacement there will, on average, be a slightly higher density of
galaxies and these being more distant will have a slightly higher
Hubble velocity.  But if one were to bin the measurements on the
intermediate scale and create peculiar velocity estimates normalising
by the local density of galaxies, then the intermediate scale motions
would no longer couple to the larger-scale density perturbations.

In redshift space distortions, large-scale peculiar motions cause a
`squishing' of the clustering pattern, enhancing the apparent
amplitude of modes with wave vector out of the plane of the sky.  But
the peculiar velocity from smaller scale motions tends to counteract
that.  In (\ref{eq:velbias}) we see that large-scale motions are also
biased.  The velocity dispersion here is a `single-galaxy' or
`1-point' statistic.  In RSD we deal with the pairwise velocity
dispersion.  At small projected separation, where both pairs most
probably reside in the same virialised halo the pairwise averaging
gives a large variance as it tends to be dominated by large clusters.
But for pairs with substantial projected separation the distribution
function is the convolution of the distribution for two places which,
as far as small-scale motions are concerned provide a velocity
dispersion very similar to that in (\ref{eq:velbias}).  Thus it would
seem that measurements of the velocity bias given by
(\ref{eq:velbias}) -- which depends to some extend on assumptions
about the galaxy density bias since $\beta = f(\Omega) / b$ -- could
provide independent constraints on the distribution of small-scale
motions that is the nuisance factor in RSD measurements.

In $\Lambda$CDM the pairwise line-of-sight velocity dispersion falls
from about 700 km/s at $\sim 1 {\rm Mpc}/h$ scales to about 500 km/s
at a few Mpc/$h$ and above (Jenkins et al.\ 1998).  This is consistent with the idea
that at small separation one is seeing pairs in the same halo, so one
is measuring the velocity dispersion of clusters weighted by the
square of the number of galaxies in those clusters, rather than weighting galaxies
equally.  The estimate from LCRS (Jing, Mo \& Borner 1997) is flat at
$\simeq$ 500 km/s out to a few Mpc (the uncertainty becoming large for
larger separation).  We interpret this to mean that at these
separations the pairwise velocity difference distribution is the
convolution of the single-particle distribution with itself so that
the appropriate value for the 1-particle RMS velocity is
$\sigma_v = 500 / \sqrt{2} \simeq 350$ km/s.   With $\beta \simeq 0.5$ (i.e.\ $b
\simeq 1$) this suggests that the bias in (\ref{eq:velbias}) would be
a $\sim$ 25\% effect for $k \sim 0.1 h / {\rm Mpc}$ (or wavelength
$\lambda \simeq 60 {\rm Mpc} / h$).

Koda et al.\ 2013 have calculated the power spectrum of peculiar
velocities in numerical cosmological simulations.  They observed a
damping at high frequencies of the velocity power as measured in
redshift space that they found could be described by a `sinc'
function: $D_u = \sin(k \sigma_K / H) / (k \sigma_K / H)$ with
$\sigma_K \simeq 1300$ km/s.  Expanding this to lowest non trivial
order would give $D_u(k) \simeq 1 - k^2 \sigma_K^2 / 3! H^2$ or
equivalently $D_u(k) = 1 - k^2 \sigma_v^2 / \beta H^2$ with $\sigma_v
\simeq 375$ km/s (with $\beta = 0.5$), in excellent agreement with our
prediction.  However, we would emphasise that this 1-particle velocity
dispersion from simulations is considerably larger than the value
seemingly preferred by the data of Willick \& Strauss (1998) who found
a ML value of $\sigma_v \simeq 140$ km/s.  

Equation (\ref{eq:velbias}) can also be used to estimate the bias of
velocities caused by errors in
photometric redshifts based on broad-band filter photometry.  But the
results, for conventional broad-band imaging surveys delivering
$\sim 5$\% fractional errors in $1+z$ are quite depressing.  
A wider range of scales can be probed
using J-PAS with its 56 narrow band filters, which is designed to
deliver much better precision of $\delta (1 + z) \simeq 0.003$.

\end{document}